\documentclass[twocolumn,aps,showpacs,preprintnumbers,amsmath,amssymb]{revtex4}

\usepackage{bm}

\begin{document}

\title{Toward unification: the dimensionless equation of motion}
\author{Lubo\v{s} Neslu\v{s}an}%
 \email{ne@ta3.sk}
\affiliation{%
 Astronomical Institute of the Slovak Academy of Sciences,\\
 05960 Tatransk\'{a} Lomnica, Slovakia
}%

\date{\today}

\begin{abstract}
   We demonstrate that if masses and charges figuring in the equation of
motion including both Newton gravitational and Coulomb electrostatic
force laws are divided by mass and charge, respectively, which are
derived using the relations contaning only the fundamental physical
and mathematical constants (like relations defining the Planck's mass,
length,  and time), then the gravitational constant and permitivity of
vacuum can be eliminated from the equation. In addition, the equation
becomes dimensionless containing only the ratios of distances,
velocities, masses, and electric charges. The ratios of masses and
charges can further be replaced with the ratios of wave-lengths or
frequencies. The corresponding equation of motion implies that the
fundamental physical constants as the gravitational constant,
permitivity of vacuum, and Planck constant are, likely, mere the
transformation constants between artificial quantities as mass and
electric charge, which were established by man to communicate some
concerning things and events in every-day life, and natural physical
quantities as wave-length or frequency of oscillations of waving
space-time.
\end{abstract}

\pacs{03.50.De, 04.20.Cv, 04.25.-g, 12.10.-g}

\keywords{equation of motion -- grand unification -- fundamental
 physical constants}

\maketitle

\section{Introduction}

   It is known that Albert Einstein during a long period of his
scientific career attempted to work out a unified theory of all forces
in the nature. He expected that such a theory would be completely
geometrical. The physical constants should be eliminated or calculated
one with the help of others.

   The goal attempted by Einstein and many his followers seems to be
achieved, at least in part, within the being-newly-born theory of the
unification based on the Maxwell
electromagnetism\footnote{arXiv:1012.5763v1 $[$physics.gen-ph$]$ 28
Dec 2010}. The outline of this theory provides a dimensionless equation
of motion of an electrically charged particles in the force field of
both gravity and electric force. Unfortunately, the theory is not yet
complete. Besides other, the theory has an ambition to include the
description of atom within the unified concept, but only a part of
energy levels in the hydrogen atom can, at this moment, be determined,
those characterized with the quantum numbers $l = n - 1$ and
$j = l + 1/2$ in the Dirac's theory. Further, the theory is expected to
provide an exact description of the mechanism of interaction, i.e. the
absorption of the wave associated with a test particle (TP) by an acting
particle (AP). At the randomly distributed phase shift between the
oscillations of the TP wave and AP itself, the model should result in
relation
\begin{equation}\label{cond}
\frac{<I_{\phi}>}{2I_{t}} = 4\pi \alpha,
\end{equation}
where $<$$I_{\phi}$$>$ is the integral characterizing the mean
absorption in unit of the maximum absorption, $I_{t} = 2/\pi$ is the
mean appearance of the TP wave in the place of AP in unit of maximum
appearance, and $\alpha$ is the fine structure constant. The preliminary
model based on a scalling law gives $<$$I_{\phi}$$> = (1 - 2/\pi )/\pi$,
which implies the calculated value of fine structure constant
$\alpha_{c} = (1 - 2/\pi )/(16\pi ) = 1/138.33$, i.e. too much (about
$1\%$) differing from the actual experimental value.

   It nevertheless appears that the dimensionless equation of motion can
also be derived regardless any new theory. In this short research note,
we describe this derivation and discuss some implications of the
dimensionless equation of motion toward the unification of gravity with
the electric interaction.

\section{Equation derived using the Planck quantities}

   Let us consider a TP in the coordinate frame in which this particle
is in rest and an AP forcing the TP to accelerate. The TP is situated in
distance $r$ from the AP. The mass and electric charge of the TP (AP)
are $m_{T}$ and $q_{T}$ ($m_{A}$ and $q_{A}$), respectively. The
acceleration of the TP, i.e. the change of its velocity $\Delta v$ per
a time interval $\Delta t$, can be calculated from its equation of
motion:
\begin{equation}\label{eqm1}
m_{T}\frac{\Delta v}{\Delta t} = -G\frac{m_{T}m_{A}}{r^{2}} +
 \frac{1}{4\pi \varepsilon_{o}}\frac{q_{T}q_{A}}{r^{2}},
\end{equation}
where $G$ is the gravitational constant and $\varepsilon_{o}$ is the
permitivity of vacuum.

   The dimensional analysis enables combining the fundamental physical
constants to obtain special length, time, and mass, which are known as
the Planck length, $L_{P}$, Planck time, $t_{P}$, and Planck mass,
$M_{P}$. Specifically, these quantities are defined by
\begin{equation}\label{LPtPMP}
L_{P} = \sqrt{\frac{G\hbar}{c^{3}}},~~
t_{P} = \frac{L_{P}}{c} = \sqrt{\frac{G\hbar}{c^{5}}},~~
M_{P} = \frac{\hbar}{cL_{P}} = \sqrt{\frac{\hbar c}{G}}
\end{equation}
In these relations, $c$ is the velocity of light and $\hbar$ is the
Planck's constant divided by $2\pi$. Further, we know that the fine
structure constant can be given as
$\alpha = q_{o}^{2}/(4\pi \varepsilon_{o}\hbar c)$ ($q_{o}$ is the
elementary electric charge), from which
$1/(4\pi \varepsilon_{o}) = \alpha \hbar c/q_{o}^{2}$. If $G$ is
expressed with the help of the Planck's mass, i.e.
$G = \hbar c/M_{P}^{2}$, and we assume the change of velocity $\Delta v$
during the Planck's time, i.e. $\Delta t = t_{P}$, then Eq.(\ref{eqm1})
can be re-written to form
\begin{equation}\label{eqm2}
m_{T}\frac{M_{P}c^{2}}{\hbar}\Delta v = -\frac{\hbar c}{r^{2}}
 \frac{m_{T}m_{A}}{M_{P}^{2}} + \frac{\alpha \hbar c}{r^{2}}
 \frac{q_{T}q_{A}}{q_{o}^{2}}.
\end{equation}

   Further, if we multiply the right-hand side of this equation with
$L_{P}^{2}M_{o}^{2}c^{2}/\hbar^{2}$ (i.e. with unity) and, then,
multiply the whole equation with $\hbar /(M_{P}c^{2})$, the equation
acquires the dimensionless form
\begin{equation}\label{eqm3}
\frac{m_{T}}{M_{P}}\frac{\Delta v}{c} = -\left( \frac{L_{P}}{r}
 \right) ^{2}\frac{m_{T}}{M_{P}}\frac{m_{A}}{M_{P}} + \alpha \left(
 \frac{L_{P}}{r}\right) ^{2} \frac{q_{T}}{q_{o}}\frac{q_{A}}{q_{o}}.
\end{equation}

\section{Usage of elementary electromass and related quantities}

   A more natural way to define the fundamental interval of mass and,
subsequently, length and time than a pure dimensional analysis is
a formal unification of Newton gravitational and Coulomb electrostatic
laws. Specifically, we define such a mass, $M_{o}$, that two particles
having this mass attract each other with the same force as particles
charged with two (positive and negative) elementary electric charges.
It means, we require the validity of
$GM_{o}^{2}/r^{2} = q_{o}^{2}/(4\pi \varepsilon_{o} r^{2})$. The latter
yields $M_{o} = q_{o}/\sqrt{4\pi \varepsilon_{o} G} = \sqrt{\alpha} M_{P}$.
We refer to this mass as the "elementary electromass", hereinafter.
The interval of length appears to be suitably defined as the Compton
wave-length corresponding to $M_{o}$ divided by the factor of $2\pi$.
In the new theory of unification, mentioned in Sect.~1, this length is
called as "interaction radius", $R_{I}$. According to our definition,
$R_{I} = \hbar /(2\pi M_{o}c)$. Finally, let us assume that the TP
accelerates during the time interval equal to $\Delta t = 2R_{I}/c =
t_{P}/(\pi \sqrt{\alpha})$. With the new set of fundamental intervals,
Eq.(\ref{eqm1}) can be re-written to the form
\begin{equation}\label{eqm4}
\frac{m_{T}}{M_{o}}\frac{\Delta v}{c} = 4\pi \alpha \left( \frac{R_{I}}
 {r}\right) ^{2} \left( -\frac{m_{T}}{M_{o}}\frac{m_{A}}{M_{o}} +
 \frac{q_{T}}{q_{o}}\frac{q_{A}}{q_{o}}\right) .
\end{equation}
Eqs.(\ref{eqm3}) and (\ref{eqm4}) are both dimensionless, because only
the dimensionless quantities and the ratios of masses, velocities,
lengths, and charges, which are of course also dimensionless, occur.

   A pure geometrical description can, in principle, contain only two
quatities: length and, if the geometry is evolving in time, the time
gradient of lenght. Eqs.(\ref{eqm3}) as well as (\ref{eqm4}) can
acquire the form containing only these quantities when we assume the
TP and AP consisting of only positively and negatively charged
elementary particles (EPs). Specifically, we assume that the TP (AP)
consists of $N_{+}$ ($n_{+}$) positively charged EPs, each with mass
$m_{+}$ and charge $+q_{o}$, and $N_{-}$ ($n_{-}$) negatively charged
EPs with mass $m_{-}$ and charge $-q_{o}$. So,
$m_{T} = N_{+}m_{+} + N_{-}m_{-}$, $m_{A} = n_{+}m_{+} + n_{-}m_{-}$,
$q_{T} = (N_{+} - N_{-})q_{o}$, and $q_{A} = (n_{+} - n_{-})q_{o}$.
Using this assumption and denotation, Eq.(\ref{eqm4}) can be written
in form
\begin{eqnarray}\label{eqm5}
\left( N_{+}\frac{m_{+}}{M_{o}} + N_{-}\frac{m_{-}}{M_{o}}\right) 
 \frac{\Delta v}{c} = 4\pi \alpha \left( \frac{R_{I}}{r}\right) ^{2} .
 \nonumber \\
 .\left[ -\left( N_{+}\frac{m_{+}}{M_{o}} + N_{-}\frac{m_{-}}{M_{o}}
 \right) \left( n_{+}\frac{m_{+}}{M_{o}} + n_{-}\frac{m_{-}}{M_{o}}
 \right) \right. + \nonumber \\
 \left. + \left( N_{+} - N_{-}\right) \left( n_{+} - n_{-} \right)
 \right] .
\end{eqnarray}

   Further, we can use the well-know de Broglie's relation between the
mass of particle and the frequency of its associated wave, $\nu$, i.e.
$mc^{2} = h\nu$ ($h$ is the original Planck's constant). If the
frequency is converted to the wave-length, $\lambda$, according to
$\nu = c/\lambda$, then mass can be given as $m = h/(c\lambda)$. We also
assign an associated wave, with wave-length $\Lambda_{o}$, to the
elementary electromass. Using these conversions and writing
corresponding subscripts in the denotation, Eq.(\ref{eqm5}) can be,
after some algebraic handling, written in another form:
\begin{eqnarray}\label{eqm6}
\left( N_{+}\frac{\Lambda_{o}}{\lambda_{+}} + N_{-}\frac{\Lambda_{o}}
 {\lambda_{-}}\right) \frac{\Delta v}{c} = 4\pi \alpha \left(
 \frac{R_{I}}{r}\right) ^{2} . \nonumber \\
 .\left[ -\left( N_{+}\frac{\Lambda_{o}}{\lambda_{+}} + N_{-}
 \frac{\Lambda_{o}}{\lambda_{-}}\right) \left( n_{+}\frac{\lambda_{o}}
 {\lambda_{+}} + n_{-}\frac{\Lambda_{o}}{\lambda_{-}}\right) + \right.
 \nonumber \\
 \left. + \left( N_{+} - N_{-}\right) \left( n_{+} - n_{-} \right)
 \right] .
\end{eqnarray}

   Ratio $\Delta v/c$ in Eq.(\ref{eqm6}) expresses a gradient of change
of the TP position given in unit of change of position of a photon
moving with the velocity of light. Besides the "gradient" of change of
the EP position (length), Eq.(\ref{eqm6}) contains only ratios of
length and numbers of EPs. So, it satisfies the requirements of pure
geometrical description. The masses and charges can be eliminated in
an analogous way also from Eq.(\ref{eqm3}).

   One could, perhaps, object that the fine structure constant,
figuring in Eq.(\ref{eqm6}), though dimensionless is still physical
constant and, therefore, this equation is not purely geometrical. This
problem can be removed assuming an alternative representation of
$\alpha$. In Sect.~1, we introduced an example, where $\alpha$ was
related, by Eq.(\ref{cond}), to geometrical aspects of the mechanism
of interaction. Accepting, e.g., this representation, $\alpha$ could be
calculated from the geometry of interaction and was not, thus, the true
physical, but rather geometrical (i.e. mathematical) constant.

\section{Some steps toward the unification}

    In the previous part, we considered the wave associated with
particle, which uses to be described by wave equation. The amplitude of
the wave, in one-dimensional case, is often obtained in form of
exponential function $\exp (\pm ikr)$, where $i$ is the unit of
imaginary numbers and $k$ is the size of wave-vector. If interaction
is assumed to be mediated by its associated wave and signal is received
by the particle at an extremely short distance ($r = R_{I}$), then the
argument $\pm ikr \rightarrow \pm ikR_{I}$. In a simple case, the size
of wave-vector can be given as the ratio of angular frequency, $\omega$,
and velocity of light, and the former can be given with the help of
mass, $m$, according the well-known de Broglie's relation
$mc^{2} = \hbar \omega$. So, the argument $\pm ikR_{I}$ can be gauged to
become $\pm im/M_{o}$. We note that $m/M_{o} \ll 1$ for all EPs,
therefore the exponential can be approximted, with a high precision, as
$\exp (\pm im/M_{o}) = 1 \pm im/M_{o}$. This amplitude can further be
gauged as $-i(1 + im_{+}/M_{o})$ when corresponding with the positively
charged EP and as $+i(1 - im_{-}/M_{o})$, i.e. as complex-number
conjugate, when corresponding to the negatively charged EP.

   In electromagnetism as well as quantum physics, the state of a system
sometimes uses to be described by the complex-valued functions, the real
parts of which correspond to the actually observed effects. If we
assume that the force between two (point-like) objects is proportional
to the appropriate sums of the above-mentioned amplitudes, which equal
to $-iN_{+}(1 + im_{+}/M_{o}) + iN_{-}(1 - im_{-}/M_{o})$ for a test
object and  $-in_{+}(1 + im_{+}/M_{o}) + in_{-}(1 - im_{-}/M_{o})$ for
an acting object, then it is possible to prove that the real part of the
product of multiplication of these two sums is identical to the form in
brackets of Eq.(\ref{eqm5}).

   Interestingly, the form in the brackets in Eq.(\ref{eqm5}) includes
not only the Coulomb electrostatic, but also the Newton gravitational
law and we just reproduced these brackets only with the amplitudes of
functions which can occur in the solution of Maxwell equations. The
gravity really seems to be comprehended by the theory of
electromagnetism. In a neutral body, for which $N_{+} = N_{-} \equiv N$,
sum $-iN(1 + im_{+}/M_{o}) +iN(1 - im_{-}/M_{o})$ is equal to
$N(m_{+} + m_{-})$. It means, the first terms ($-iN$ and $+iN$) of the
exponentials, expanded to power series of argument, vanish, but the
second terms ($Nm_{+}/M_{o}$ and $Nm_{-}/M_{o}$) survive. And, notice
that the result $N(m_{+} + m_{-})$ is, in fact, the mass of the neutral
body consisting of $N$ positively charged EPs, with mass $m_{+}$, and
$N$ negatively charged EPs, with mass $m_{-}$.

\section{Consequent representation of physical constants}

   In Sects.~2 and 3, we could see that if masses and charges figuring
in the equation of motion including both Newton gravitational and
Coulomb electrostatic force laws are divided by mass and charge,
respectively, which are derived using the relations contaning only the
fundamental physical and mathematical constants, then the gravitational
constant and permitivity of vacuum can be eliminated from the equation.

   The true meaning of the constants $G$, $\epsilon_{o}$, and $h$ (or
$\hbar$) seems to be indicated by Eq.(\ref{eqm6}) in which also the
masses and charges are eliminated. In this equation, we meet only the
distances and wave-lengths (which could be converted, eventually, to
frequencies), besides the gradient of the change of position. Seeing
this equation, it seems that the nature "knows" only the curved and
oscillating space-time and change in position of wave sources.

   Actually, the concept of mass evolved from the concept of weight of
material bodies in the Earth's gravity and concept of electric charge
was established as a certain analogy to the concept of mass. The
weight/mass was not initially any physical quantity, but people
established this concept in dawn of ages to mutually communicate, also
quantitatively, the measure of some goods or effects of the Earth's
attraction of material things. When the force laws were later created
using the concepts of mass and/or charge, there had to be established
some "transformation constants" between the artificially defined human
quantities (mass, charge) and natural physical quantities (wave-lengths
or frequencies of oscillations). If human quantities are not used in
the formulation of force laws, no transformation constants, as seen in
Eq.(\ref{eqm6}), are needed. Just this seems to be the role of such the
fundamental physical constants as gravitational constant, permitivity of
vacuum, and the Planck constant.

\begin{acknowledgments}
   This work was supported, in part, by the VEGA, Slovak Grant Agency
for Science (grant No. 0011).
\end{acknowledgments}


\end{document}